\documentclass[12pt,a4paper]{article}

\usepackage{comment}
\usepackage{tikz}

\usepackage{caption}
\usepackage{mathtools}

\usepackage[mathscr]{euscript}

\usepackage{slashed}

\usepackage{amsmath}

\date{}

\title{\textbf{
Bopp-Podolsky Scalar Electrodynamics Propagators and Energy-Momentum Tensor in Covariant and Light-Front Coordinates
}}
\author{ \textbf{
Ilane Gomes Oliveira$^{a}$, Jorge Henrique Sales$^{a}$ and
Ronaldo Thibes$^{b}$}
\\\\
\textit{$^{a}$\small{Universidade Estadual de Santa Cruz}}\\
\textit{\small{Ilh\'eus - BA, Brazil}}\\
\textit{$^{b}$\small{Universidade Estadual do Sudoeste da Bahia}}\\
\textit{\small{Rodovia BR 415, Km 03, S/N -- Itapetinga, Bahia, Brazil}}
 }

\usepackage{graphicx}
\usepackage{amssymb,amsfonts,amssymb,amsthm}
\usepackage{amsmath}

\begin{document}

\maketitle

\abstract{
We consider the interaction between
a charged scalar boson and the Bopp-Podolsky gauge fields.
The Bopp-Podolsky (BP) electrodynamics possesses both massive and massless propagation modes for the photon, while preserving gauge invariance.
We obtain the propagator
of all fields present in the model for the higher-order generalizations of the linear covariant,
light-front and doubly transverse light-front gauges.
Although
BP's original model is described by a
higher-order derivatives Lagrangian, it is possible to work with
an equivalent reduced-order version by means of the introduction of an auxiliary vector field.
We compute the gauge-invariant improved energy-momentum tensor for the full reduced-order interacting BP model.
Besides the more traditional front-form view, we also discuss the light-front perspective in both versions of the model.
Within a Lagrangian framework approach we
maintain explicit covariance at all steps and show that the field propagators, as well as the energy-momentum tensor, can be cast into a light-front closed form using specific properties of general coordinate transformations.
}

\section{Introduction}\label{int}
The physical mass spectrum of a given quantum field theory is one of its inherent characteristic fingerprints, connecting its theoretical predictions with concrete experimental data.  In general, null masses signal the presence of important symmetries and vice-versa.
As an important example, the gauge symmetry of quantum electrodynamics (QED), implemented in the usual way, renders the photon massless.
The Bopp-Podolsky (BP) model \cite{Bopp, Podolsky:1942zz} provides an interesting way to generate mass to gauge fields without breaking gauge invariance.
Introducing higher-order derivative terms in the Lagrangian, the BP model generalizes ordinary electrodynamics maintaining gauge invariance, linearity and enhancing convergence.
The initial historical motivation for the BP model \cite{Bopp, Podolsky:1942zz, Podolsky:1944zz, Podolsky:1948} was to get rid of some of the embarrassing infinities in electrodynamics by means of using higher-order derivatives resulting in higher momentum powers upon Fourier transformation, at the cost of introducing an additional open parameter $a$.
For
instance, the resulting BP electrostatic potential $V(r)$ is given as a function of the distance $r$ by
\begin{equation}\label{BPpotential} 
V(r) = \frac{1-e^{-r/a}}{r}
\,,
\end{equation}
amounting to a generalization of the traditional Coulomb's potential
clearly finite at the origin\footnote{In the sense that $\lim_{r\rightarrow 0^+} V(r) = 1/a$.}.  Nevertheless, in view of the large developments of quantum field theory along the last century, 
the infinities were tamed by renormalization and
quantum electrodynamics has become one of the most firmly well established physical frameworks, both on conceptual and experimental grounds.
In this context, in modern terms, we have two chiefly possible interpretations for the BP model.  The first one explores the mentioned fact that a $1/a$ new massive mode can be generated without breaking gauge invariance, providing thus a tool for constructing possible extensions of the standard model or effective field theories containing massive vector fields.
The second interpretation considers the real variable $a$ in equation (\ref{BPpotential}) as a non-physical Pauli-Villars regulating parameter \cite{Slavnov1, Slavnov2, Stoilov, Ji:2019phv}.
In this latter case, due to the higher-order derivatives present in the model, one is able to control the divergences as in the Pauli-Villars regularization scheme and follow the usual renormalization program as can be seen in detail in reference \cite{Ji:2019phv}.

Parallel to these facts, the formalism of light-front (LF) dynamics has gradually established itself as a serious important general quantization framework for quantum field theory, particularly well-suitable for dealing with bound states in quantum chromodynamics (QCD).
In 1949, Paul Dirac published an important well-known paper \cite{Dirac:1949cp}
in which, in order to harmonize relativity and quantum theories, three different
forms of Hamiltonian dynamics were proposed.  Those came to be known
as the point-form, light-front-form and the usual instant-form, depending
on how one chooses the initial conditions as well as the dynamical evolution
parameter.
In the present paper we shall be concerned with the instant and light-front forms.
While in
the usual instant-form the physical time defines dynamics,
the light-front-form
relies on the so-called light-front evolution parameter, in our context defined as $x^+\equiv(x^0+x^3)/\sqrt{2}$.
Although $x^+$ is not a physical time, it stems from a genuine invertible change of variables and can
be picked up as the evolution parameter for the equations of motion in the Hamiltonian formalism \cite{Dirac:1949cp, Bakker:2000mn}.
Actually, the use of light-front coordinates dates back to d'Alembert who made
use of them to obtain the general solution for the initial value problem of
the bidimensional wave-equation in a closed form \cite{DAlembert:1747}.
Naturally, in reference \cite{DAlembert:1747}, d'Alembert does not use neither quantum mechanics nor relativity, but rather shows that a simple change of variables to LF coordinates is sufficient to immediately produce the corresponding partial differential equation general solution in an elegant and neat way.

The first concrete applications of LF techniques to particle physics originated from the peculiar idea of an infinite momentum limit Lorentz frame
\cite{Fubini:1964boa, Weinberg:1966jm, Susskind:1967rg} which leads to considerable simplifications in the Feynman diagrams of perturbation theory.
The connection was done when 
Bardakci and Halpern showed that the so-called infinitum momentum frame could be interpreted as a LF coordinates change \cite{Bardakci:1969dv} allowing Kogut and Soper to establish the initial foundations of quantum electrodynamics in the infinite momentum frame 
\cite{Kogut:1969xa}.
Since then, LF quantization methods have been consistently applied to QCD both in the perturbative and non-perturbative regimes as can be seen for instance in the representative reference works \cite{Lepage:1980fj, Brodsky:1980ex, Wilson:1994fk, Ji:1995kf, Cao:1997hw} or comprehensive reviews
\cite{Brodsky:1997de, Heinzl:2000ht, Brodsky:1991ir, Brodsky:2003vn, Bakker:2013cea, Hiller:2016itl}.
It is worth mentioning that
connections between the LF and instant forms through an interesting interpolating scheme appeared in the initial works by
Hornbostel \cite{Hornbostel:1991qj} and C. R. Ji and C. Mitchell \cite{Ji:2001xd} and have been 
recently more deeply explored in references
\cite{Ji:2012ux, Ji:2014vha, Ji:2018pic, Ji:2018cqg}.

With the enormous conceptual and technical development
over the last years of
quantum field theory as a whole,
there exist currently various nice working, rigorous and powerful quantization methods. In particular, many
of them do not rely directly on the more traditional Hamiltonian approach, as for instance integral
functional quantization methods, with the additional benefit of being explicitly relativistic
covariant throughout all quantization steps.
Still one is always allowed to use LF coordinates and transit from one specific approach to another.  In the present paper we propose to investigate
the scalar Bopp-Podolsky (BP) model along these lines.
More precisely, starting from the total action for the scalar
BP model, we perform a change of coordinates obtaining the
field propagators and energy-momentum tensor in Minkowsky space both in Lorentz covariant and LF coordinates.
Since those are connected by a linear transformation, a matrix approach is suitable -- and in fact very handy -- permitting one to write explicitly the propagators and energy-momentum tensor in LF coordinates in a very straightforward manner.  As a result we obtain a consistent field theory whose all features, such as propagators, energy-momentum tensor and Green functions, can be directly described in LF and compared to the traditional Hamiltonian LF approach.

The current text is organized as follows.  In section {\bf 2} below, we formally introduce the BP scalar electrodynamics by coupling the BP gauge field to a charged scalar field.  We discuss further the gauge invariance of the resulting model, its field equations and obtain the full energy-momentum tensor density.
In section {\bf 3} we turn to LF and compute the corresponding energy-momentum tensor LF components and gauge field propagators for three distinct gauge fixings.
In section {\bf 4} we consider the recently introduced reduced-order  BP model \cite{Thibes:2016ivt} interacting with the charged bosonic field obtaining the LF energy-momentum tensor.
We conclude in section {\bf 5} with a few final remarks.

\section{Bopp-Podolsky Scalar Electrodynamics}
The BP scalar electrodynamics consists of
the interaction between a charged bosonic field
$\phi$ and a Bopp-Podolsky gauge field $A_\mu$, with coupling constant $e$, given
by the Lagrangian density\footnote{The Lorentz
indexes run from $0$ to $3$ in Minkowsky space with metric convention $\eta^{\mu\nu} = \mbox{diag}(1,-1,-1,-1)$.}
\begin{equation}\label{Lint}
{\cal L}_{int} = ieA_\mu
\left[
\phi\partial^\mu\phi^* - \phi^*\partial^\mu\phi
\right] + e^2A^2|\phi|^2
\,,
\end{equation}
with $|\phi|^2\equiv \phi^*\phi$.
The corresponding dynamics
for the complex scalar $\phi$ and real
spin one $A_\mu$ fields are given respectively by
the Lagrangians
\begin{equation}\label{Lphi}
{\cal L}_\phi = \partial_\mu \phi^* \partial^\mu\phi - m^2|\phi|^2\,,
\end{equation}
and
\begin{equation}\label{LA}
{\cal L}_A =
-\frac{1}{4}F_{\mu\nu}F^{\mu\nu}
 +\frac{
 a^2
 }{2}\partial_\nu F^{\mu\nu}\partial^\rho F_{\mu\rho}
\,.
\end{equation}
The first one, ${\cal L}_\phi$ in (\ref{Lphi}), describes a usual complex Klein-Gordon field, while ${\cal L}_A$ stands for the Bopp-Podolsky higher-derivatives
model
\cite{Bopp, Podolsky:1942zz,
Podolsky:1944zz, Podolsky:1948}
and leads to a neat generalization of Maxwell's electrodynamics 
-- modern recent reviews can be seen for instance in
\cite{Ji:2019phv, Thibes:2016ivt, Lazar}.
As mentioned in the Introduction, the BP model comes with an extra open parameter $a$ which can be seen in equation (\ref{LA}) multiplying the term containing the higher-order derivatives.
This extra length-dimensional parameter $a$, as will be seen below in the corresponding equations of motion, is responsible for a new massive mode
for the photon
without breaking gauge invariance.
In fact,
in equation (\ref{LA}), $F_{\mu\nu}$ denotes the ordinary field strength tensor given
by
\begin{equation}\label{Fst}
F_{\mu\nu} = \partial_\mu A_\nu - \partial_\nu A_\mu
\end{equation}
which is trivially gauge-invariant under
\begin{equation}\label{gt1}
A_\mu\rightarrow A_\mu - \partial_\mu\Lambda
\end{equation}
for a given arbitrary space-time dependent function $\Lambda$.
Thus,
although (\ref{LA}) depends on higher-order derivatives of the gauge field
$A_\mu$, since they come in the exact combination (\ref{Fst}), the Bopp-Podolsky Lagrangian ${\cal L}_A$
is still gauge-invariant.
The parameter $a$ also provides a natural length scale for the model and can be shown to regularize the usual Coulomb potential to the corresponding BP potential (\ref{BPpotential}) as can be seen in \cite{Ji:2019phv}.
Regarding the scalar field $\phi$, the Lagrangian ${\cal L}_\phi$ alone is not invariant
under
\begin{equation}\label{gt2}
\phi \rightarrow e^{ie\Lambda}\phi
\,,
\end{equation}
transforming rather as
\begin{equation}
{\cal L}_\phi \rightarrow
{\cal L}_\phi + ie \partial_\mu \Lambda
\left[\phi \partial^\mu \phi^* - \phi^*\partial^\mu\phi \right]
+e^2 |\phi|^2 \partial_\mu\Lambda\partial^\mu\Lambda
\,.
\end{equation}
This variation though, is exactly canceled by the one coming from the interaction Lagrangian in equation (\ref{Lint}).  Actually, ${\cal L}_{int}$ can be thought as
resulting from substituting the ordinary derivative in (\ref{Lphi}) by the covariant one
\begin{equation}
D_\mu = \partial_\mu  +ie A_\mu
\,
\end{equation}
and the sum ${\cal L}_{\phi}+{\cal L}_{int}$ is invariant under the combined transformations
(\ref{gt1}) and (\ref{gt2}) for arbitrary $\Lambda$.

By demanding stationarity of the total gauge-invariant action
\begin{equation}\label{S}
S=\int d^4x\,\left[ {\cal L}_\phi + {\cal L}_A + {\cal L}_{int} \right]
\end{equation}
with respect to arbitrary variations of the
scalar and gauge
fields $\phi$ and $A_\mu$, we obtain the following system of coupled differential equations
\begin{equation}\label{EL1}
(\Box+m^2)\phi = -ieA_\mu\partial^\mu\phi -ie\partial^\mu(\phi A_\mu ) + e^2 A^2 \phi\,,
\end{equation}
\begin{equation}\label{EL1b}
(\Box+m^2)\phi^* = ieA_\mu\partial^\mu\phi^* + ie\partial^\mu(\phi^* A_\mu) + e^2 A^2 \phi^*
\end{equation}
and
\begin{equation}\label{EL2}
(1+a^2\square)\partial_\nu F^{\mu\nu}= ie(\phi\partial^\mu\phi^*-\phi^*\partial^\mu\phi)
+2e^2 A^\mu|\phi|^2
\,.
\end{equation}
As usual,
the box symbol $\square$ denotes the d'Alembertian
second-order differential
operator, here given by $\square \equiv \partial_\mu\partial^\mu$.
Note that (\ref{EL1}) and (\ref{EL1b}) are the complex conjugate of each other, while (\ref{EL2}) may be equivalently rewritten
as
\begin{equation}\label{EL3}
  \left(
  \partial^\mu\partial^\nu
  - \Box\eta^{\mu\nu}
  \right)
  \left(1+a^2\Box\right)
  A_\nu= ie
  \left[\phi \partial^\mu \phi^* - \phi^*\partial^\mu\phi \right]
+2e^2 A^\mu|\phi|^2
\,.
\end{equation}
The RHS of either (\ref{EL2}) or (\ref{EL3}) can be interpreted as a source term for the gauge field $A_\mu$ in a fourth-order partial differential equation.  It is clear then from (\ref{EL3}) that $A_\mu$ has two propagating modes, namely a massless and a $1/a$ massive one.

By considering the Noether current associated to space-time translations, the canonical energy-momentum tensor for the total action (\ref{S}) can be directly computed as
\begin{equation}\label{canonicalEMT}
{\Theta}^{\mu\nu} = {\Theta}^{\mu\nu}_\phi + {\Theta}^{\mu\nu}_A + {\Theta}^{\mu\nu}_{int}
\end{equation}
with
\begin{equation}\label{Thetaphi}
{\Theta}^{\mu\nu}_\phi \equiv \partial^\mu \phi^* \partial^\nu\phi
+ \partial^\mu \phi \partial^\nu \phi^*
-\eta^{\mu\nu}\partial_\rho\phi^*\partial^\rho\phi +m^2\eta^{\mu\nu}|\phi|^2
\,,
\end{equation}
\begin{equation}\label{ThetaA}
\begin{array}{rcl}
{\Theta}^{\mu\nu}_A &\equiv&\left[F^{\rho\mu}-\displaystyle\frac{a^2}{2}\partial^\rho\partial_\lambda F^{\mu\lambda} + a^2\partial^\mu\partial_\lambda F^{\rho\lambda}
\right]\partial^\nu A_\rho
\displaystyle+\frac{a^2}{2}\partial_\rho F^{\lambda\rho}\partial_\lambda\partial^\nu A^\mu
\\&&
\displaystyle+\frac{a^2}{2}\partial_\rho F^{\mu\rho}\partial_\lambda\partial^\nu A^\lambda
-a^2\partial_\rho F^{\lambda\rho}\partial^\mu\partial^\nu A_\lambda
+\eta^{\mu\nu}
\left[
\frac{1}{4}F_{\lambda\rho}F^{\lambda\rho}
 -\displaystyle\frac{
 a^2
 }{2}\partial_\lambda F^{\sigma\lambda}\partial^\rho F_{\sigma\rho}
\right]
\end{array}
\end{equation}
and
\begin{equation}\label{Thetaint}
{\Theta}^{\mu\nu}_{int} \equiv
ie\left[
\eta^{\mu\alpha}\eta^{\nu\beta}-\eta^{\mu\nu}\eta^{\alpha\beta}
\right]
A_\alpha
\left[
\phi\partial_\beta\phi^*-\phi^*\partial_\beta\phi
\right]
-e^2\eta^{\mu\nu}A^2|\phi|^2
\,.
\end{equation}
A straightforward calculation of the divergence with respect to the first index $\mu$ in
(\ref{canonicalEMT}) leads to
\begin{eqnarray}
\partial_\mu \Theta^{\mu\nu} &=&
\phantom{\frac{1}{2}\!\!\!\!\!}\left[
(\Box + m^2)\phi^*-ie\partial_\mu(A^\mu\phi^*)-e^2A^2\phi^*-ie A_\mu \partial^\mu \phi^*
\right]\partial^\nu\phi
\nonumber\\&&
+\phantom{\frac{1}{2}\!\!\!\!\!}\left[
(\Box + m^2)\phi+ie\partial_\mu(A^\mu\phi)-e^2A^2\phi+ie A_\mu \partial^\mu \phi
\right]\partial^\nu\phi^*
\nonumber\\&&
+\left[-ie(\phi\partial^\mu\phi^*-\phi^*\partial^\mu\phi)
-2e^2A^\mu|\phi|^2+(1+a^2\square)\partial_\rho F^{\mu\rho}
\right]\partial^\nu A_\mu
\end{eqnarray}
which shows explicitly the consistent conservation of $\Theta^{\mu\nu}$ modulo the equations of motion (\ref{EL1}), (\ref{EL1b}) and (\ref{EL2}).
However, $\Theta^{\mu\nu}$ is neither gauge-invariant nor symmetric.  This can be fixed by constructing the so-called improved energy-momentum tensor \cite{Belinfante}
\begin{equation}\label{Tmunu}
{\cal T}^{\mu\nu} = \Theta^{\mu\nu}_\phi + {\cal T}^{\mu\nu}_A + {\cal T}^{\mu\nu}_{int}
\end{equation}
with
\begin{eqnarray}\label{TAmunu}
{\cal T}^{\mu\nu}_A &\equiv&
F^\mu_{\;\;\,\rho}F^{\rho\nu}
+a^2\left(
F^{\mu\rho} \square F_\rho^{\;\;\,\nu}+F^{\nu\rho} \square F_\rho^{\;\;\,\mu}
-\partial_\lambda F^{\mu\lambda}\partial_\rho F^{\nu\rho}
\right)
\nonumber\\&&
+\frac{1}{2}\eta^{\mu\nu}\left(
\frac{1}{2}F^{\rho\lambda}F_{\rho\lambda} + a^2 F^{\rho\lambda}\square F_{\rho\lambda}
+a^2\partial_\lambda F^{\sigma\lambda} \partial^\rho F_{\sigma\rho}
\right)
\end{eqnarray}
and
\begin{equation}\label{Tintmunu}
{\cal T}^{\mu\nu}_{int} \equiv
ieA_\alpha\left(
\eta^{\mu\alpha}\eta^{\nu\beta}
+\eta^{\mu\beta}\eta^{\nu\alpha}
-\eta^{\mu\nu}\eta^{\alpha\beta}
\right)
\left[
\phi\partial_\beta\phi^*-\phi^*\partial_\beta\phi
-ieA_\beta|\phi|^2
\right]
\,,
\end{equation}
which now may be checked to be indeed gauge-invariant, symmetric and divergenceless.
This concludes our brief presentation of the model in its explicit Lorentz covariant form.  We shall come to its LF dynamics in the next section.
\section{Light-Front Dynamics and Propagators}
In the previous section we have discussed the BP scalar electrodynamics using
Lorentz covariant space-time
coordinates $x^\mu$, $\mu=0,1,2,3$, with the index $0$ denoting physical time.
Inspired by d'Alembert well-known solution for the 2-dimensional wave equation \cite{DAlembert:1747}, in order to connect our results to the vast LF quantization literature \cite{Bakker:2013cea, Ji:2013tea, Bertin:2009gs, Bertin:2017orz},
we now introduce the following light-front coordinates
\begin{equation}\label{LFt}
x^{\pm}=\frac{\sqrt{2}}{2}(x^0\pm x^3)
\,.
\end{equation}
The variables $x^\pm$ represent the longitudinal LF coordinates, whereas the remaining ones are usually packed into a perpendicular two-dimensional vector ${\mathbf x^\bot}\equiv (x^1,x^2)$. 
Note that (\ref{LFt}) is not a Lorentz transformation as it does not
preserve the internal product in Minkowski space and therefore does not connect two inertial frames.
Nevertheless, starting from the Minkowski metric
tensor $\eta^{\mu\nu} = \mbox{diag}(1,-1,-1,-1)$, we may define
a transformed LF metric as
\begin{equation}\label{etaab}
\eta^{ab}=\left(
\begin{array}{cccc}
0&0&0&1\\0&-1&0&0\\0&0&-1&0\\1&0&0&0
\end{array}
\right)=\eta_{ab}
\end{equation}
with the Latin indexes running within $a,b = +,1,2,-$.
Then it follows immediately that we may summarize the LF coordinates transformation
as
\begin{equation}\label{25}
x^a = T^a_{\;\;\mu}x^\mu
\end{equation}
and
\begin{equation}\label{26}
x_a = T_a^{\;\;\mu}x_\mu
\,,
\end{equation}
with the four-by-four invertible matrices
\begin{equation}\label{Tamu}
T^a_{\;\;\mu}\equiv
\left(
\begin{array}{cccc}
\sqrt{2}/2&0&0&\sqrt{2}/2\\
0&1&0&0\\
0&0&1&0\\
\sqrt{2}/2&0&0&-\sqrt{2}/2
\end{array}
\right)
\mbox{~~~and~~~}
T_a^{\;\;\mu}\equiv
\left(
\begin{array}{cccc}
\sqrt{2}/2&0&0&\sqrt{2}/2\\
0&1&0&0\\
0&0&1&0\\
\sqrt{2}/2&0&0&-\sqrt{2}/2
\end{array}
\right)
\end{equation}
satisfying
\begin{equation}\label{Tproperties}
T^a_{\;\;\mu} = \eta^{ab}\eta_{\mu\nu}T_b^{\;\;\nu}
\,,~~~~~~~~
T^a_{\;\;\mu}\,T_a^{\;\;\nu}=\delta_\mu^\nu\,,~~~~~~~~\mbox{ and }~~~~~~~~
T_a^{\;\;\mu}\,T^b_{\;\;\mu} =\delta_a^b
\,.
\end{equation}
It is also clear from (\ref{Tamu}) that the coordinate transformation (\ref{LFt}) has unity Jacobian, which renders the space-time volume element invariant.

The LF transformation, initially defined for the space-time coordinates (\ref{25}) and (\ref{26}), carry on naturally to arbitrary Lorentz tensors.  In particular, if $V^\mu$ and $W^\mu$ denote two given Lorentz vectors, we have the general dot product identity
\begin{equation}\label{dotproduct}
V_a\,W^a=(T_a^{\;\;\mu}\, T^a_{\;\;\nu})\,V_\mu\,W^\nu
=V_\mu\,W^\mu
\,.
\end{equation}
This implies that the Lagrangian terms present in the total action (\ref{S}) can be cast into LF form
as
\begin{equation}\label{LphiLF}
{\cal L}_\phi = \partial_a \phi^* \partial^a \phi - m^2|\phi|^2
\,,
\end{equation}
\begin{equation}\label{LALF}
{\cal L}_A =
-\frac{1}{4}F_{ab}F^{ab}
 +\frac{
 a^2
 }{2}\partial_b F^{ab}\partial^c F_{ac}
\,,
\end{equation}
and
\begin{equation}\label{LintLF}
{\cal L}_{int} = ieA_a
\left[
\phi\partial^a \phi^* - \phi^*\partial^a \phi
\right] + e^2A^2|\phi|^2
\,.
\end{equation}
By using the LF metric (\ref{etaab}), the index sums in (\ref{LphiLF}), (\ref{LALF}) and (\ref{LintLF}) can be fully split in LF coordinates.  In fact,
if the indexes $i,j,k$ represent the two remaining perpendicular spatial coordinates which are not affected by (\ref{LFt}), i.e., $i,j,k = 1,2$, we have
\begin{equation}
{\cal L}_{\phi} = \partial_+\phi^*\partial_-\phi
+\partial_-\phi^*\partial_+\phi
-\partial_i\phi^*\partial_i\phi-m^2|\phi|^2
\,,
\end{equation}
\begin{eqnarray}
{\cal L}_A &=& 
\frac{1}{2}F_{+-}F_{+-}
+F_{+i}F_{-i}
-\frac{1}{4}F_{ij}F_{ij}
+a^2\left(\partial_+F_{+-}+\partial_iF_{i+}\right)
\left(\partial_-F_{-+}+\partial_jF_{j-}\right)
\nonumber\\&&
-\frac{a^2}{2}
\left(
\partial_+F_{-i}+\partial_-F_{+i}+\partial_jF_{ij}
\right)
\left(
\partial_+F_{-i}+\partial_-F_{+i}+\partial_kF_{ik}
\right)
\,,
\end{eqnarray}
and
\begin{eqnarray}
{\cal L}_{int}&=&ieA_+\left[ \phi\partial_-\phi^* - \phi^*\partial_-\phi \right]
+ieA_-\left[ \phi\partial_+\phi^*-\phi^*\partial_+\phi \right] 
\nonumber\\&&\phantom{\frac{1}{2}\!\!\!\!\!}
-ieA_i\left[ \phi\partial_i\phi^*-\phi^*\partial_i\phi \right]
+e^2A^2|\phi|^2
\,.
\end{eqnarray}

In this way we can easily convert any result from covariant to LF coordinates.  To illustrate further the idea, we
write next the improved energy-momentum tensor in the light-front.
By using the conversion matrices (\ref{Tamu}) along with their properties (\ref{Tproperties}), we obtain from (\ref{Tmunu}), (\ref{TAmunu}) and (\ref{Tintmunu}) the relations
\begin{eqnarray}
{\cal T}^{++}&=&
2\partial_-\phi^*\partial_-\phi
+2ieA_-(\phi\partial_-\phi^*-\phi^*\partial_-\phi -ieA_-|\phi|^2)
\nonumber\phantom{\frac{1}{2}\!\!\!\!\!}\\&&
+\,F_{-i}F_{-i}+2a^2F_{-i}\square F_{-i} - a^2(\partial_-F_{-+}+\partial_iF_{i-})^2
\,,
\end{eqnarray}
\begin{eqnarray}
{\cal T}^{+-}&=&\partial_i\phi^*\partial_i\phi+m^2|\phi|^2
+\frac{1}{2}F_{+-}F_{+-}+\frac{1}{4}F_{ij}F_{ij}
+a^2F_{+-}\square F_{+-}
\nonumber\\&&
+\frac{a^2}{2}F_{ij}\square F_{ij}
+ieA_i\left(
\phi\partial_i\phi^*-\phi^*\partial_i\phi-ieA_i|\phi|^2
\right)
\nonumber\phantom{\frac{1}{2}\!\!\!\!\!}\\&&
-\frac{a^2}{2}(\partial_+F_{-i}+\partial_-F_{+i}+\partial_j F_{ij})
(\partial_+F_{-i}+\partial_-F_{+i}+\partial_k F_{ik})
\,,
\end{eqnarray}
\begin{eqnarray}
{\cal T}^{--}&=&
2\partial_+\phi^*\partial_+\phi
+2ieA_+(\phi\partial_+\phi^*-\phi^*\partial_+\phi-ieA_+|\phi|^2)
\nonumber\phantom{\frac{1}{2}\!\!\!\!\!}\\&&
+\,F_{+i}F_{+i}+2a^2F_{+i}\square F_{+i} - a^2(\partial_+F_{+-}+\partial_iF_{i+})^2
\,,
\end{eqnarray}
\begin{eqnarray}
{\cal T}^{\pm i}&=&-\partial_\mp\phi^*\partial_i\phi - \partial_\mp\phi\partial_i\phi^*
+F_{\pm\mp}F_{\mp i} + F_{\mp j}F_{ji}
+a^2(F_{\pm\mp}\square F_{\mp i} + F_{\mp j}\square F_{ji})
\nonumber\phantom{\frac{1}{2}\!\!\!\!\!}\\&&
+a^2(F_{\mp i}\square F_{\pm\mp}+F_{ij}\square F_{j\mp}
)
+a^2(\partial_\mp F_{\pm\mp} + \partial_j F_{\mp j})
(\partial_{\pm} F_{\mp i} + \partial_\mp F_{\pm i} + \partial_k F_{ik})
\nonumber\phantom{\frac{1}{2}\!\!\!\!\!}\\&&
-ieA_\mp(\phi\partial_i\phi^*-\phi^*\partial_i\phi-ieA_i|\phi|^2)
-ieA_i(\phi\partial_\mp\phi^*-\phi^*\partial_\mp\phi-ieA_\mp|\phi|^2)\phantom{\frac{1}{2}\!\!\!\!\!}
\,,
\end{eqnarray}
and
\begin{eqnarray}\label{Tij}
{\cal T}^{ij}&=&
\partial_i\phi^*\partial_j\phi+\partial_i\phi\partial_j\phi^*
+{ie}(\delta_{ik}\delta_{jl}+\delta_{il}\delta_{jk}-\delta_{ij}\delta_{kl})
A_k\left[\phi\partial_l\phi^*-\phi^*\partial_l\phi-ieA_l|\phi|^2\right]
\nonumber\phantom{\frac{1}{2}\!\!\!\!\!}\\&&
+\delta_{ij}\left[\phantom{\frac{1}{2}\!\!\!\!\!}
ieA_+(\phi\partial_-\phi^*-\phi^*\partial_-\phi-ieA_-|\phi|^2)
+ieA_-(\phi\partial_+\phi^*-\phi^*\partial_+\phi-ieA_+|\phi|^2)
\right.
\nonumber\\&&
\left.+\frac{1}{2}F_{+-}F_{+-}+F_{+k}F_{-k}-\frac{1}{4}F_{kl}F_{kl}+\partial_+\phi^*\partial_-\phi+\partial_-\phi^*\partial_+\phi-\partial_k\phi^*\partial_k\phi-m^2|\phi|^2\right]
\nonumber\\&&\phantom{\frac{1}{2}\!\!\!\!\!}
+F_{i+}F_{-j}+F_{i-}F_{+j}+F_{ik}F_{jk}
+a^2\tilde{\cal T}^{ij}
\,,
\end{eqnarray}
with\footnote{The indexes $l, m, n$ in equations (\ref{Tij}) and (\ref{Ttij}), similarly to $i, j, k$, also vary along the perpendicular direction assuming the values $1$ and $2$.}
\begin{eqnarray}\label{Ttij}
\tilde{\cal T}^{ij}&\equiv&F_{i+}\Box F_{-j}+F_{i-}\Box F_{+j}
-F_{ik}\Box F_{kj}+F_{j-}\Box F_{+i}+F_{j+}\Box F_{-i}
-F_{jk}\Box F_{ki}
\nonumber\\&&\phantom{\frac{1}{2}\!\!\!\!\!}
-(\partial_+ F_{i-}+\partial_- F_{i+}+\partial_k F_{ki})
(\partial_+ F_{j-}+\partial_- F_{j+}+\partial_l F_{lj})
\nonumber\\&&
-\delta_{ij}\left[
F_{+-}\Box F_{-+}+F_{k-}\Box F_{+k}+F_{k+}\Box F_{-k}
+\frac{1}{2}F_{kl}\Box F_{kl}
\right.\nonumber\\&&\phantom{\frac{1}{2}\!\!\!\!\!}\left.
+(\partial_+ F_{+-} +\partial_k F_{k+})(\partial_- F_{-+} +\partial_l F_{l-})
\right.\nonumber\\&&\left.
-\frac{1}{2}
(\partial_+ F_{-k}+\partial_- F_{+k}+\partial_m F_{km})
(\partial_+ F_{-k}+\partial_- F_{+k}+\partial_n F_{kn})
\right]
\,.
\end{eqnarray}

In order to study the model perturbatively, one has to 
deal with the field propagators.  Our next task then is to compute the field propagators in the light-front.  
This can be done by inverting the operators acting on the quadratic field terms of the Lagrangian densities (\ref{LphiLF}) and (\ref{LALF}) and performing a Fourier transformation to momentum space $k^a$.
Since ${\cal L}_\phi$ is not gauge-invariant by itself, the propagator for the scalar field $(\phi,\phi^*)$ can be readily computed in momentum space to be given by
\begin{equation}\label{Pphi}
P_{(\phi,\phi^*)} =
\left(
\begin{matrix}
0&\frac{1}{k^2-m^2}\\\frac{1}{k^2-m^2}&0
\end{matrix}
\right)
\,.
\end{equation}
As $k^2$ is invariant, due to (\ref{dotproduct}), the scalar field propagator (\ref{Pphi}) has the same form either in covariant or LF coordinates.
Concerning the vectorial sector, associated to the BP field $A_a$,
we need first to fix the gauge freedom. In the following, we consider three main gauge-fixing possibilities, namely,
the higher-order generalizations of the linear covariant, the non-covariant LF and the doubly transverse LF gauges.

The generalization of the linear covariant gauge for the BP model can be implemented by inserting the $\xi$-dependent gauge-fixing term
\begin{equation}\label{GF1}
{\cal L}_{GF1}=-\frac{1}{2\xi}\partial_a A^a (1+a^2\square)\partial_b A^b
\,,
\end{equation}
into the total action (\ref{S}).  In (\ref{GF1}), $\xi$ denotes a free gauge parameter.
This permits us to invert the BP field quadratic terms in the gauge-fixed action leading to the $\xi$-dependent expressions
\begin{equation}\label{44}
P_{++}^{(1)}=\frac{(\xi-1)k_+^2}{(1-a^2k^2)k^4}\,,\,\,\,\,\,\,\,\,\,\,\,\,\,\,P_{--}^{(1)}=\frac{(\xi-1)k_-^2}{(1-a^2k^2)k^4}\,,
\end{equation}
\begin{equation}\label{45}
P_{+-}^{(1)}=\frac{(\xi+1)k_+k_--k_ik_i}{(1-a^2k^2)k^4}\,,\,\,\,\,\,\,
\,\,\,\,\,\,
P_{+i}^{(1)}=\frac{(\xi-1)k_+k_i}{(1-a^2k^2)k^4}
\,,
\end{equation}
and
\begin{equation}\label{46}
P_{ij}^{(1)} =
\frac{-\delta_{ij}k^2+(\xi-1)k_ik_j}{(1-a^2k^2)k^4}
\end{equation}
for the vector field propagator in momentum space.
Note that the BP parameter $a$ considerably enhances the convergence of the model by assuring an
extra $k^2$ in the denominators of all propagators.
Besides the usual $k^2=0$ pole, it is also clear from (\ref{44}) to (\ref{46}) that the BP field propagator exhibits an additional one at $k^2=1/a^2$, confirming once more its extra massive propagating mode.
The limits $\xi\rightarrow0$, $\xi\rightarrow1$, $\xi\rightarrow3$ and $\xi\rightarrow\infty$ in the expressions above  correspond to the natural higher-order generalizations of, respectively, the Landau, Feynman-t'Hooft, Fried-Yennie and unitary gauges.

As a second gauge-fixing possibility, we investigate next the non-covariant choice
\begin{equation}\label{GF2}
{\cal L}_{GF2}=-\frac{1}{\alpha}A_+(1+a^2\square)A_+
\end{equation}
containing the free gauge parameter $\alpha$.  This choice corresponds to the subsidiary gauge condition
\begin{equation}
(1+a^2\square)A_+ = 0
\,,
\end{equation}
which generalizes the usual light-front gauge $A_+=0$ by including an $a$-dependent piece proportional to the d'Alembertian.  
The addition of the space-time integral of (\ref{GF2}) to the total action also renders the BP quadratic term invertible and leads, in the limit $\alpha\rightarrow0$, to
\begin{equation}\label{P2++}
P_{++}^{(2)}=P_{+-}^{(2)}=P_{+i}^{(2)}=0
\,,
\end{equation}
\begin{equation}\label{P2--}
P_{--}^{(2)}=\frac{-2k_-}{k^2(1-a^2k^2)k_+}
\,,\,\,\,\,\,\,\,\,\,\,\,\,
P_{-i}^{(2)}=\frac{-k_i}{k^2(1-a^2k^2)k_+}
\,,
\end{equation}
and
\begin{equation}\label{P2ij}
P_{ij}^{(2)}=\frac{-\delta_{ij}}{k^2(1-a^2k^2)}
\,.
\end{equation}
Finally, as a third possibility, we consider the non-covariant doubly transverse LF gauge \cite{Srivastava:2000cf, Suzuki:2003jz} which can be described by
\begin{equation}\label{GF3}
{\cal L}_{GF3}=-\frac{\sqrt{2}}{\beta}A_+(1+a^2\square)(\partial \cdot A )
\,.
\end{equation}
In this last case, the inclusion of (\ref{GF3}) produces a non-singular total action whose vector propagators, in the limit $\beta\rightarrow0$, coincide with the previous ones in equations (\ref{P2++}) to (\ref{P2ij}) with the sole exception of the new
\begin{equation}
P_{--}^{(3)} = - \frac{k_ik_i}{k^2(1-a^2k^2)k_+^2}
\,.
\end{equation}
This last result assures the transversality of the propagator with respect to the momentum in this specific gauge, i.e.,
\begin{equation}
k^a P^{(3)}_{ab}=0\,.
\end{equation}
Note that this property does not hold for neither of the two previous cases.

\section{The Reduced-Order Model}
As we have seen, the generalized BP photon corresponding to the action
(\ref{S}) carries an extra massive mode.  However, the ordinary photon massless mode is still
present as is clear from equation (\ref{EL3})
and from the fact that all propagators for the gauge field obtained in the last section have two poles, at $k^2=0$ and $k^2=1/a^2$.
In order to decouple these two
massive and massless modes and associate them to distinct physical fields
we may work with the reduced-order Lagrangian \cite{Thibes:2016ivt}
\begin{equation}\label{LB}
{\cal L}_{B}
= -\frac{1}{4}F_{\mu\nu} F^{\mu\nu}
-\frac{a^2}{2}B_\mu B^\mu
+a^2\partial_\mu B_\nu F^{\mu\nu}
\,,
\end{equation}
where $B_\mu$ denotes an auxiliary massive vector field.
As an important bonus, the introduction of $B_\mu$ permits to reduce the
order of space-time derivatives.  It has been shown in \cite{Thibes:2016ivt}
that the reduced-order Lagrangian (\ref{LB}) is equivalent to (\ref{LA})
both at classical and quantum levels.
In this last section we compute the LF energy-momentum tensor for a corresponding
reduced-order BP scalar electrodynamics.

By coupling (\ref{LB}) to the matter bosonic field, the total action for the reduced-order model reads
\begin{equation}\label{Sr}
S_{(r)} = \int\,d^4x\left\{
{\cal L}_\phi + {\cal L}_{B} + {\cal L}_{int} 
\right\}
\,,
\end{equation}
being gauge-invariant under the combined gauge transformations
\begin{equation}
A_\mu\rightarrow A_\mu-\partial_\mu\Lambda\,,\,\,\,\,\,\,
\phi\rightarrow e^{ie\Lambda}\phi\,,\,\,\,\,\,\,
B_\mu\rightarrow B_\mu\,.
\end{equation}
By imposing stationarity of (\ref{Sr}) under arbitrary field variations, we obtain
the field equations
\begin{equation}
(\Box+m^2)\phi = -ieA_\mu\partial^\mu\phi -ie\partial^\mu(\phi A_\mu ) + e^2 A^2 \phi\,,
\end{equation}
\begin{equation}
(\Box+m^2)\phi^* = ieA_\mu\partial^\mu\phi^* + ie\partial^\mu(\phi^* A_\mu) + e^2 A^2 \phi^*
\,,
\end{equation}
\begin{equation}\label{Awave}
(\partial^\mu\partial^\nu-\square\eta^{\mu\nu})A_\nu =  B^\mu
\phantom{\frac{1}{2}\!\!\!\!\!}
\,,
\end{equation}
and
\begin{equation}\label{Bwave}
(\partial^\mu\partial^\nu-\square\eta^{\mu\nu})
(A_\nu - a^2 B_\nu)
=
 ie
  \left[\phi \partial^\mu \phi^* - \phi^*\partial^\mu\phi \right]
+2e^2 A^\mu|\phi|^2
\,.
\end{equation}
It is interesting to note that the $B_\mu$ field is divergenceless.
Indeed applying $\partial_\mu$ to both sides of (\ref{Awave}) we obtain
\begin{equation}
\partial_\mu B^\mu = 0
\,.
\end{equation}
Furthermore, using this last result and substituting (\ref{Awave}) into (\ref{Bwave}), we see that $B_\mu$ satisfies a $1/a$ mass Klein-Gordon equation with source, namely,
\begin{equation}
(1+a^2\Box)B_\mu
=
 ie
  \left[\phi \partial^\mu \phi^* - \phi^*\partial^\mu\phi \right]
+2e^2 A^\mu|\phi|^2
\,.
\end{equation}

The canonical energy-momentum tensor can be computed directly from (\ref{Sr}) as
\begin{equation}\label{Thetar}
\Theta^{\mu\nu}_{(r)} = \Theta^{\mu\nu}_\phi + \Theta^{\mu\nu}_{B} + \Theta^{\mu\nu}_{int}
\end{equation}
with $\Theta^{\mu\nu}_\phi$ and $\Theta^{\mu\nu}_{int}$ given by equations (\ref{Thetaphi}) and (\ref{Thetaint}),
and
\begin{eqnarray}
\Theta^{\mu\nu}_{B} &=&
F^{\rho\mu}\partial^\nu A_\rho
+a^2(\partial^\mu B^\lambda - \partial^\lambda B^\mu)\partial^\nu A_\lambda
+a^2F^{\mu\lambda}\partial^\nu B_\lambda
\nonumber\\&&
+\frac{1}{4}\eta^{\mu\nu}F_{\lambda\rho}F^{\lambda\rho}
+{a^2}\eta^{\mu\nu}\left(\frac{1}{2}B_\lambda B^\lambda -\partial_\lambda B_\rho F^{\lambda\rho}\right)
\,.
\end{eqnarray}
By applying the Belinfante procedure, as described in \cite{Belinfante}, the canonical energy-momentum tensor (\ref{Thetar}) can be upgraded to the symmetric gauge-invariant version
\begin{eqnarray}
{\cal T}^{\mu\nu}_{(r)}&=&\Theta^{\mu\nu}_\phi+{\cal T}^{\mu\nu}_{int}
+F^\mu_{\;\;\,\rho}F^{\rho\nu}+\frac{1}{4}\eta^{\mu\nu}F^{\rho\lambda}F_{\rho\lambda}
+{a^2}\eta^{\mu\nu}\left(\frac{1}{2}B^\lambda B_\lambda-\partial_\lambda B_\rho F^{\lambda\rho}\right)
\nonumber\\&&
+a^2F^\mu_{\;\;\,\lambda}(\partial^\nu B^\lambda - \partial^\lambda B^\nu)
+a^2F^\nu_{\;\;\,\lambda}(\partial^\mu B^\lambda - \partial^\lambda B^\mu)
-a^2B^\mu B^\nu
\,.
\end{eqnarray}
In order to convert this result to LF, we may apply the technique discussed in the last section.  In fact, using the transformation matrices (\ref{Tamu}), we define
\begin{equation}\label{Tabr}
{\cal T}^{ab}_{(r)} = T^a_{\;\;\mu} T^b_{\;\;\nu} {\cal T}^{\mu\nu}_{(r)}
\end{equation}
as the LF reduced-order energy-momentum tensor.
Considering $a,b=+,1,2,-$ for LF indexes and $i,j,k,l=1,2$ for spatial indexes, equation (\ref{Tabr}) leads explicitly to
\begin{eqnarray}
{\cal T}^{++}_{(r)}&=&
2\partial_-\phi^*\partial_-\phi
+2ieA_-(\phi\partial_-\phi^*-\phi^*\partial_-\phi -ieA_-|\phi|^2)
\nonumber\phantom{\frac{1}{2}\!\!\!\!\!}\\&&
+\,F_{-i}F_{-i}-2a^2F_{-i}(\partial_- B_i - \partial_i B_-) - a^2 B_-^2
\,,
\end{eqnarray}
\begin{eqnarray}
{\cal T}^{+-}_{(r)}&=&\partial_i\phi^*\partial_i\phi+m^2|\phi|^2
+\frac{1}{2}F_{+-}F_{+-}+\frac{1}{4}F_{ij}F_{ij}
-\frac{a^2}{2}B_i B_i - \frac{a^2}{2}F_{ij}\partial_i B_j
\nonumber\\&&
+ieA_i\left(
\phi\partial_i\phi^*-\phi^*\partial_i\phi-ieA_i|\phi|^2
\right)
+a^2F_{+-}(\partial_-B_+-\partial_+B_-)
\,,
\end{eqnarray}
\begin{eqnarray}
{\cal T}^{--}_{(r)}&=&
2\partial_+\phi^*\partial_+\phi
+2ieA_+(\phi\partial_+\phi^*-\phi^*\partial_+\phi-ieA_+|\phi|^2)
\nonumber\phantom{\frac{1}{2}\!\!\!\!\!}\\&&
+\,F_{+i}F_{+i}
-2a^2F_{+i}(\partial_+ B_i - \partial_i B_+)
-a^2 B_+^2
\,,
\end{eqnarray}
\begin{eqnarray}
{\cal T}^{\pm i}_{(r)}&=&-\partial_\mp\phi^*\partial_i\phi - \partial_\mp\phi\partial_i\phi^*
+F_{\pm\mp}F_{\mp i} + F_{\mp j}F_{ji}
+a^2F_{\mp j}(\partial_i B_j - \partial_j B_i)
\nonumber\phantom{\frac{1}{2}\!\!\!\!\!}\\&&
-ieA_\mp(\phi\partial_i\phi^*-\phi^*\partial_i\phi-ieA_i|\phi|^2)
-ieA_i(\phi\partial_\mp\phi^*-\phi^*\partial_\mp\phi-ieA_\mp|\phi|^2)\phantom{\frac{1}{2}\!\!\!\!\!}
\nonumber\\&&
+a^2 F_{i\mp}(\partial_{\pm} B_{\mp} -\partial_{\mp} B_{\pm})
+a^2F_{ij}(\partial_{\mp}B_j - \partial_j B_{\mp})\phantom{\frac{1}{2}\!\!\!\!\!}
\nonumber\\&&
+a^2F_{\mp\pm}(\partial_{\mp} B_i - \partial_i B_{\mp})
+a^2B_{\mp} B_i
\,,
\end{eqnarray}
and
\begin{eqnarray} 
{\cal T}^{ij}_{(r)}&=&
\partial_i\phi^*\partial_j\phi+\partial_i\phi\partial_j\phi^*
+{ie}(\delta_{ik}\delta_{jl}+\delta_{il}\delta_{jk}-\delta_{ij}\delta_{kl})
A_k\left[\phi\partial_l\phi^*-\phi^*\partial_l\phi-ieA_l|\phi|^2\right]
\nonumber\phantom{\frac{1}{2}\!\!\!\!\!}\\&&
+\delta_{ij}\left[\phantom{\frac{1}{2}\!\!\!\!\!}
ieA_+(\phi\partial_-\phi^*-\phi^*\partial_-\phi-ieA_-|\phi|^2)
+ieA_-(\phi\partial_+\phi^*-\phi^*\partial_+\phi-ieA_+|\phi|^2)
\right.
\nonumber\\&&
\left.+\frac{1}{2}F_{+-}F_{+-}+F_{+k}F_{-k}-\frac{1}{4}F_{kl}F_{kl}+\partial_+\phi^*\partial_-\phi+\partial_-\phi^*\partial_+\phi-\partial_k\phi^*\partial_k\phi-m^2|\phi|^2\right]
\nonumber\\&&\phantom{\frac{1}{2}\!\!\!\!\!}
+F_{i+}F_{-j}+F_{i-}F_{+j}+F_{ik}F_{jk}
+a^2{\cal T}^{ij}_B
\,,
\end{eqnarray}
with
\begin{eqnarray}
{\cal T}^{ij}_B &\equiv& F_{i+}(\partial_j B_- -\partial_- B_j) + F_{i-}(\partial_j B_+ -\partial_+ B_j)
-F_{ik}(\partial_j B_k - \partial_k B_j)
\nonumber\\&&\phantom{\frac{1}{2}\!\!\!\!\!}
+F_{j+}(\partial_i B_- - \partial_- B_i)
+F_{j-}(\partial_i B_+ - \partial_+ B_i)
-F_{jk}(\partial_i B_k - \partial_k B_i)
\nonumber\\&&
-B_iB_j-\delta_{ij}\left[
B_+ B_- - \frac{1}{2}B_kB_k +F_{+-}(\partial_+ B_- - \partial_- B_+)
\right.\nonumber\\&&\left.
+F_{k+}(\partial_k B_- - \partial_- B_k)
+F_{k-}(\partial_k B_+ - \partial_+ B_k)
-F_{kl}\partial_k B_l
\phantom{\frac{1}{2}\!\!\!\!\!}\right]
\,.
\end{eqnarray}
Thus we have obtained the full expressions for the LF energy-momentum tensor components describing the interacting reduced-order BP model.

\section{Conclusion}
We have discussed the coupling of the Bopp-Podolsky gauge field to a charged bosonic matter field in covariant and LF coordinates
and obtained the energy-momentum tensor and propagators in the light-front.
The canonical energy-momentum tensor following directly from the model defining action, although being conserved modulo field equations and thus consistent with Noether's theorem, is neither symmetric nor gauge-invariant.  Fixing that issue, we were able to obtain an equivalent improved symmetric energy-momentum tensor describing the full interacting model and enjoying gauge-symmetry.
We have then computed all LF components for this improved higher-order derivatives gauge-invariant energy-momentum tensor.
Concerning the gauge field propagators,
three gauge-fixings were analyzed, generalizing the linear covariant, light-front and doubly transverse light-front gauges for the higher-order derivatives case.  In all of them it was possible to identify two poles in the propagators, corresponding to a massless and a $1/a$ massive modes for the vector field.  Specific values of the gauge-fixing parameter in the Lagrangian lead to thorough higher-order generalizations of the Landau, t'Hoof and Fried-Yennie gauges in all of which a common term proportional to $k^2(1-a^2k^2)$ shows up in the denominator, considerably enhancing the convergence of the Feynmann integrals.  This fact permitted to clearly show that the generalized Landau gauge propagator still enjoys the transversality property of ordinary electrodynammics.  We have also obtained the gauge field propagators for a generalized LF gauge and shown how a higher-order derivatives non-covariant gauge fixing choice can produce a doubly transverse propagator.
Although neatly generalizing electrodynamics maintaining gauge invariance, the BP model has a downside regarding the use of higher-order derivatives.  Addressing that issue, we have shown that it is possible to work with an equivalent reduced-order model, at the cost of introducing an extra massive vector field.  We have pursued a comparative analysis of this reduced-order model, extending the previous work \cite{Thibes:2016ivt}, promoting the interaction with the complex scalar field and computing the corresponding improved gauge-invariant energy-momentum tensor and obtaining its components in the LF.

\subsubsection*{Acknowledgments:}
J. H. S. was supported by Conselho Nacional de Desenvolvimento Cient\'\i tico e Tecnol\'ogico, 315519/2018-5; Funda\c{c}\~ao de Amparo \`a Pesquisa do Estado da Bahia, PIE 0013/2016; and Coordena\c{c}\~ao de Aperfeiçoamento de Pessoal de N\'\i vel Superior. I. G. O. was supported by Funda\c{c}\~ao de Amparo \`a Pesquisa do Estado da Bahia, BOL0037/2017.

\end{document}